\documentclass[aoas,MSNbibl,nameyear,dvips]{arximspdf}

\doi{10.1214/11-AOAS529}
\volume{5}
\issue{4}
\pubyear{2011}
\firstpage{2265}
\lastpage{2265}

\begin{document}
\begin{frontmatter}

\title{Special section on modern multivariate analysis}

\begin{aug}
\author[A]{\fnms{Karen} \snm{Kafadar}\corref{}\ead[label=e1]{kkafadar@indiana.edu}}
\affiliation{Indiana University}
\address[A]{Department of Statistics
\\
Indiana University
\\
Bloomington, Indiana 47408-3825
\\
USA\\
\printead{e1}} 

\end{aug}




\end{frontmatter}
A critically challenging problem facing statisticians is the
identification of a suitable framework which consolidates data of
various types, from different sources, and across different time frames
or scales (many of which can be missing), and from which appropriate
analysis and subsequent inference
can proceed.

Special Section Guest Editor Susan Holmes has assembled four articles
that demonstrate the power of the duality diagram approach for analyzing
data of different formats. The first article by De la Cruz and Holmes
introduces the duality diagram and provides examples of familiar
multivariate approaches (e.g., principal components, correspondence
analysis) that fit into this framework. In the second article, Dray and
Jombart use this approach both to understand covariation structures and
to identify
spatial patterns in sociological data; this spatial application (crimes
in France) requires the incorporation of spatial constraints into the
framework. Thioulouse in the third article considers ecological data,
which arise as sets of matrices for different species and different time
points, in search of ecological changes in relationships between species
and the environment. In the fourth article, Purdom develops an approach
based on the duality diagram to combine genomic data with network
information.

We hope that these articles will lead to the identification of further
problems where the power of the duality diagram approach can be realized
to deepen the analyses of data from multiple sources, types, and
dimensions. We invite future submissions in this area that further
develop the ideas in these articles and illustrate their advantages on
real data.

\printaddresses

\end{document}